\documentclass[preprint,showpacs,preprintnumbers,amsmath,amssymb]{revtex4}

\usepackage{graphicx}% Include figure files
\usepackage{dcolumn}% Align table columns on decimal point
\usepackage{bm}% bold math
\usepackage{epsfig}

\def\be{\begin{equation}}
\def\ee{\end{equation}}
\def\bea{\begin{eqnarray}}
\def\eea{\end{eqnarray}}

\begin{document}
\title{Can we see naked singularities?}
\author{Shrirang S. Deshingkar}
\affiliation{Harish-Chandra Research Institute, Chhatanag Road, Jhunsi, Allahabad,
211019}

\begin{abstract}
We study singularities which can form in a spherically symmetric gravitational collapse of a 
general matter field obeying weak energy condition. We show that no energy can reach an outside observer from a null naked singularity. That means they will not be a serious threat to
the Cosmic Censorship Conjecture (CCC). For the timelike naked singularities, where only the central shell gets singular, the redshift is always finite and they can in principle, carry energy to a faraway observer. 
Hence for proving or disproving  CCC  the study of timelike 
naked singularities will be more important.
Our results are very general and are independent of initial data and the form of the matter.
\end{abstract}

\pacs{0.42.Dw, 0.42.Jb}

\maketitle

%\section{Introduction}

%\label{s-intro}

	In the late stages of stellar evolution, when the star burns out all its nuclear fuel
and the mass of the remnant is large enough, no known force in nature can support the core against its own gravity. In such a case we expect  continued gravitational collapse leading to a singularity. Similarly one expects there will be very small mass primordial black holes
which form in the early universe and also super-massive black holes at the galactic centers and they will also contain a singularity. The singularity theorems~\cite{HawkEllis}  guarantee formation of a singularity if there is a trapped surface. But singularities could also form without formation of
a trapped surface. In such cases information can go out from such a (naked) singularity and there can be a breakdown of predictability. We  do not know what laws of physics will apply at the singularity and typically we would like to avoid such a situation. Penrose~\cite{Pen1} proposed CCC for the same purpose. 
CCC  demands that the singularities which form in a
gravitational collapse should never be visible to an outside observer or 
they should be hidden inside a horizon. 
Proving or disproving CCC is one of the most 
important open problems in general relativity and black hole physics.  
There are many specific examples of naked singularities
as well as many cases directly or indirectly supporting
the conjecture~\cite{HaradaICgc,TBLall1, PsjPrd93, 
ShrirLum1, ShrirTime, NolanNrng,  Pfluid,  MagliTan2all, 
HaradaTimeFam, Vaidya,   OtherMagliExact, PsjCmp94, ChrisScalar}. 
Most of these studies are based on showing the existence of the 
outgoing radial null geodesics (ORNGs) from the singularity. But even if 
such geodesics come out,  whether one can actually observe them is currently  an open question. 
Most of these studies are for specific form of matter and initial data hence
cannot prove or disprove CCC immediately as formulation of CCC will need 
certain general form of matter and initial data.

In this work, we show that {\it null naked singularity} which forms 
in a spherically symmetric collapse of a general matter field   
{\it can never  be  observed}.  We conclude this  by showing that at the 
most, along one singular geodesic  the redshift remains finite, while along 
all other (infinite family of) singular geodesics the redshift diverges. 
Earlier Deshingkar et al.~\cite{ShrirLum1} had shown a similar result for 
the Tolman-Bondi-Lema$\rm \hat i$tre (TBL) dust model.
For the null singularities, this gives a strong evidence in support of CCC,  
as even though geodesics  come out from the singularity we will not be able to 
see them as no energy or information will reach an observer
from such a singularity. For the singularity  to be observed the redshift should
be finite for finite duration. We need a wavepacket to carry energy not just one or two  wavefronts.   
We also show that the redshift for {\it timelike naked singularities} 
is always finite and they {\it could be visible} to an outside observer.
Our conclusions are based on few natural assumptions and are independent 
of  the initial data and the form of the
matter (apart from it being a type I matter field ~\cite{HawkEllis}). This means study of timelike singularities
will be crucial in proving or disproving CCC. In this paper we give the basic idea and a brief argument for the proof, the details can be found elsewhere~\cite{GenRed2}.

	The metric for a general spherically symmetric
spacetime in comoving coordinates can be  written as
\be
ds^2= -e^{2 \nu} dt^2 + {e^{2\psi}}dr^2 + R^2d\Omega^2,
\label{eq:met1}
\ee
where $d\Omega^2 = d\theta^2 +\sin^2{\theta} d\phi^2$, is the line element
of two-sphere. $\nu$, $\psi$ and $R$ (also called the area radius)
are functions of $t$ and $r$. All the metric functions are
assumed to be $ C^2$ differentiable (except at the singularity).  
The matter is assumed to be any type I matter field obeying the weak energy condition.
For such matter the stress energy tensor 
$T^{a}_{b}$ has diagonal form in these coordinates and is written as, 
$
T^t_t=-\rho,\quad T^r_r=p_1,\quad T^{\theta}_{\theta}=p_2=
T^{\phi}_{\phi}=p_3,\quad T^t_r=T^r_t=0 .
$
One can write down all the Einstein equations~\cite{PsjCmp94} and knowing the 
equation of state one can devise a scheme to solve them in a self 
consistent manner.  But we only  need couple of Einstein's equations for our 
purpose and they are written as,
\be
T^t_t=-\rho=-\frac {F'} {k_0R^2R'},
\label{eq:en1}
\ee
\be
H \equiv e^{-2\nu}\dot R^2 = f + \frac{F}{R},
\label{eq:en3}
\ee
where we have defined, $f(t,r)=e^{-2\psi}(R')^2 -1$  and 
$F=F(t,r)$ is an arbitrary functions of $t$ and $r$. 
In our notation prime and over-dot denote partial derivatives with respect 
to $r$ and $t$ respectively. 
For dust and perfect fluid collapse, $F$ is the mass function.
For dust $f =f(r)$ is  the energy function.  
The $R(t,r) =0$ represents formation of a shell focusing singularity and we 
concentrate on these singularities. We assume there are no shell crossings near the
central singularity. Though the shell crossing
singularities may arise, they are believed to be weak singularities through
which extension of spacetime is expected to be  possible and hence they are not considered important
for cosmic censorship conjecture.  As there are no shell crossings ($R'>0$) near 
the center, weak energy condition implies $\rho$ is  positive i.e. $F'$ and $F$ are positive.

We just
assume that singularity forms in the gravitational collapse and perform the analysis of geodesics near the singularity. 
We will  study ORNGs, but the analysis can be easily generalized to nonradial as well as timelike geodesics. We are studying collapse i.e. $\dot R \le 0$ solution. Along ORNGs we can write, 
\be 
\frac{dR}{dr} = {\left (  1- \frac{\sqrt{f+F/R}}{\sqrt{1+f}} \right )} R'.
\label{eq:dRdr}
\ee
and the geodesic tangent vectors $K^t = \frac{dt}{d\lambda}$ ($\lambda$ is an affine parameter) 
has to satisfy the differential equation for geodesics, which on integration gives,
\be
 K^t =    \frac{ k_0 \vert {\dot R} \vert }{\sqrt{f+F/R}}
  e^{
 \int \frac{1}{\sqrt{1+f}} (\frac{\partial}{\partial r} \sqrt{f + F/R}) dr},
\label{eq:P}
\ee
where $k_0$ is constant of integration related to energy.
All the equations so far are exact and general.

From Eq.~(\ref{eq:dRdr}) we can see that, at $R=F$, the area radius along the ORNGs starts 
decreasing with increasing $r$ and that represents the apparent horizon. Typically 
we expect $F$ to be nonzero when $r > 0$. This implies that the non-central shell focusing singularity 
will always be covered as seen from 
Eq.~(\ref{eq:dRdr}). Therefore we only need to analyze the $r=0$ shell focusing singularity to check
for its visibility.

Eq.~(\ref{eq:en3}) can be written as,
$
\dot R^2 = f e^{2\nu} + \frac{F e^{2\nu}}{R} \equiv f_r +\frac{F_r}{R},
$
where in the last part we have substituted $f_r= f e^{2\nu}$ and $F_r= F e^{2\nu}$. First we check, the 
restrictions $f_r$ and $F_r$ are required to satisfy for regularity on 
an initial surface  and get a local 
solution of $R$
near the central singularity. Then we use it to see whether geodesics can come out from the singularity and calculate $K^t$, redshift and luminosity. We rescale the time coordinates to set $e^{2\nu} = 1$
at $r=0$. 

In general if we do the scaling such that on a nonsingular surface near the center (or 
just before the central singularity is formed) $R = r$, then for the density to
be finite on that surface we need to the lowest order $F_r \propto r^3$ and $f_r \propto r^2$. If the 
collapse is monotonous (or at least locally monotonous) then $R(t,r)$ can be inverted to get $t(r,R)$.
We assume $f_r$ and $F_r$
to be expandable in terms of power series in $r$, $R$ around $(r=0, R=0)$. The series can 
have positive as well as negative powers. 
In general 
$F_r = F_1(r) + (r -R)F_2(r)/(R^{b_2} + h(r))^{g_1} + ... $ with our scaling. 
We can have similar kind of expansion for $f_r$.  On a nonsingular surface $  \dot R^2 \propto r^2$ 
i.e. $R \propto r$ (else there is a contradiction). Any terms with positive power of $R$ 
in the numerator will be unimportant at the singularity 
as $R$ goes to zero faster there.  In term like $R^{b_2} + h(r)$ 
both parts have to retain their dominance then $R$ can have two possible forms.

(1) $R^{b_2} \propto h(r)$ +  $ q(t,r)$, where $q(t,r)/h(r) \rightarrow 0$
as $r \rightarrow 0$.  
In this case  collapse has to stop at  radius $R = h^{1/b_2}$ and there 
will be rebounce.  This case leads to a timelike singularity or a 
regular solution as  discussed later.

(2)  Other possibility is when $R^{b_2} \propto h(r) (t_0(r) -t)^{b_3}$. If on a non-singular surface both 
$h(r)$ and $R^{b_2}$ have equal power of $r$ then on singular surface $R^{b_2}$
term will be unimportant. If on singular surface both terms have same 
power of $r$ then $h(r)$ will be unimportant before the singular surface.  
This  leads to black holes or null naked singularities as seen below.

We can have combination of terms with $R$ and $r$ many of them giving 
same power of r on a $t= \rm constant$ non-singular or singular surface. 
But the leading order behavior in these cases remains similar to the case where there is only one term of that order. 
Hence the cases we discuss below  essentially cover  the most general kind of behavior one can get.

In the case when there are only two terms in the $\dot R^2$ equation, of 
the form $\frac{F_c(r)}{R^{b_2}}$, 
near the center integrating it we  get a hypergeometric function. As long as there is no bounce,
it can be expanded near center as some $\rm constant $ plus higher power terms in $r$ and $R$ .
Even for more general form of functions with many terms we get similar 
behavior  and we can write  near the center,
\be
R^{\alpha} \approx r^{\alpha_1} ( t_0(r) -t) \equiv Q(r) (t_0(r) -t),
\label{eq:Rform1}
\ee
where, $\alpha$ and $\alpha_1$ are constants. As such even intuitively we would expect $R$ to have above
form near the center when there is continued gravitational collapse without 
a rebounce. To satisfy Einstein 
equation~(\ref{eq:en3})  we need that till we reach the singular surface ($t =t_0(0))$,  $Q(r)$ is proportional to $r$.  The scaling $R \propto r$ has to change on singular surface ($t=t_0(0)$) for the singularity to form at the center and if different shells get singular at different times then the singular time for them
can be written as $t_0(r)$ and so $R$ near the center will have form given in Eq.~(\ref{eq:Rform1}).
Differentiating Eq.~(\ref{eq:Rform1}) we obtain,
\bea
R' \approx  r^{\beta -1} ( \eta X  + \zeta  X^{1 - \alpha}),
\label{eq:R'centre}
\eea
where $\eta \equiv \frac{rQ'}{\alpha Q} $, $X \equiv R/ r^\beta$ and typically we choose $\beta$ such that
$
\zeta \equiv \frac {Q(r) t'_0(r)} {\alpha  r^{\alpha \beta -1}} 
$
is finite. If we have geodesics coming out of the  singularity then along singular ORNGs,
\be
X_0 \equiv \frac{dR}{dr^\beta} = \lim_{r\rightarrow 0}{\left (  1
- \frac{\sqrt{f+F/R}}{\sqrt{1+f}} \right )} \frac{R'}{\beta r^{\beta -1}} \equiv U(X_0,0) \nonumber
\ee
or
\be
U(X,0) -X \equiv V(X) =0,
\label{eq:root}
\ee
where $U(X,r^\beta) = dR/dr^\beta$ along the geodesics.
If this equation has a real positive root $X_0$ with $\beta>1$ then the 
singularity is naked. 
This equation has either two real positive roots or no real positive root. 

Now to check whether a family of geodesics come out along a given direction, we write the
area radius along the geodesic as
$
	R \approx X_0 r^\beta + w(r),
$
where $w(r)$ is a function through which behavior of the family comes out. 
For the family to exist $w(r)$ should go to zero faster than $r^\beta$. 
Substituting this in the null geodesic
equation~(\ref{eq:root}) near the singularity we get,
\be
\frac{dw(r)}{dr} \approx  \frac{w(r)}{r} {\left(    1 + {\frac {dV(X)}{dX}}\mid_{X=X_0} \right)}.
\label{eq:w}
\ee
In  V(X), the coefficient of highest power of X is negative. Hence along the larger root
$\frac {dV(X)}{dX}$ is negative while it is positive along the smaller root. Integrating
Eq.~(\ref{eq:w}) we see that
along the larger root $w(r)$ does not go to zero as $r$ goes to zero and
there is only one geodesic coming out of this direction. While along smaller root 
$w(r)= D e^{\int \frac{1}{r} (1+ {\frac {dV(X)}{dX}}\mid_{X=X_0})dr}$
goes to zero and  there is an infinite family of geodesics (labeled by D) coming out of the singularity 
along this direction~\cite{ShrirTime, PsjPrd93}.

If $\beta$ is such that at the center $F/r^\beta =0$ for finite $\zeta $ then along the larger root  
$X_0 = (\zeta /(\beta - \eta))^{1/\alpha}$. While along the smaller root~\cite{ShrirQuasi,ShrirTime} 
$ R \approx F(1+ \frac{2 (1+f)}{\zeta}\frac{dF}{dr} 
\left ( \frac{F}{r^\beta} \right )^{\alpha-1})$. An infinite family of singular 
geodesic come out along this direction. From Eq. (\ref{eq:P}), we get 
that at the central singularity $K^t$ remains finite
along the larger root and it blows up as a negative exponential power of $r$ 
along the smaller root direction~\cite{ShrirLum1}.
This exponential divergence is not very surprising as these geodesic stay very close to apparent horizon near the
center.
We  get similar behavior for $K^t$ when $f_r$ term dominates near singularity and blows
up near the center for $F_r <R$.

If the value of $\beta$ for finite $\zeta $ is such that $ F/R$ is finite and  $F/R<1$  then 
from Eq.(\ref{eq:P}, \ref{eq:root}), we get a power law 
divergence of $K^t$~\cite{ShrirLum1} along all the root 
directions.
For all the cases discussed above, doing similar calculations for ingoing radial null geodesics one can see that there is only one
such geodesic terminating at the central singularity. That means the singularity is a null naked singularity.

Now we consider the other case, where only the center gets singular and the collapse stops and may be rest of the shells eventually bounce off. In such case we get a timelike naked singularity as there is no trapped surface. In this case $\dot R$ goes to zero before the apparent horizon i.e.
$f +F/R$ goes to zero for some values of $R$ even for non-central shells. In which case in the neighborhood of the bounce we can write
$
\dot R^2 \approx \frac{{b^2(r)}(R -a(r))^{2\alpha}}{R^{2 b_1}}.
$
Integrating this we again get hypergeometric function  but the collapse stops at $R=a(r)$ i.e. $ R = -F/f$. 
If $a(r) \propto r$ then there is a bounce before singularity formation.
If we have root(s) to Eq.(\ref{eq:root}) with $R \gg a$ then along those root(s) earlier analysis applies.
If not, we have to expand the hypergeometric function around $fR/F =-1$ to check the nature of the singularity,
instead of around $R=0$ as earlier.
Near $R=a$, integrating  Eq.~(\ref{eq:en3}) we get,
\bea
R & \approx a + (\frac {(\alpha -1) b}{a^{b_1}} (t - t_0(r)))^{-1/(\alpha -1)}  & \rm for \;  \alpha >1, \nonumber \\
& \approx  a + t_0(r) e^{ -t{ b}/{a^{b_1}} } &  \rm for \; \alpha =1, \nonumber \\
& \approx  a +  (\frac {(1 -\alpha ) b}{a^{b_1}} |t - t_0(r)|)^{1/(1 -\alpha)}  &  \rm for \; \alpha < 1.
\label{eq:Rbounce}
\eea
In the first case one needs $(a^{\alpha -1 -b_1} /b)^{1/(\alpha-1)} $ going to zero and in the second 
$a^{b_1}/b$ going to zero  as $r \rightarrow 0$ else our assumption of
being close to $R=a(r)$ breaks and the case comes under the previous 
category if singularity forms. In the present case only center can get 
singular and the $r \ne 0$ shell take infinite amount of proper time to 
reach the $r=a$  surface and the singularity is timelike naked singularity. 
From Eq.~(\ref{eq:dRdr}) and Eq.~(\ref{eq:root}) we can have geodesics 
coming out of this direction with behaviors like
$R(r) \approx a(r) + D ({b}/{a^{b_1}} )^{-1/(\alpha -1)}$ for the first 
case and $R(r) \approx a(r) + e^{ -D { b}/{a^{b_1}}}$ for the second case, 
where $D$ is a constant of integration
labeling the infinite family  of geodesics.
Harada et al.~\cite{HaradaTimeFam} had obtained  a exponential family 
in the case of HIN  solution for counter-rotating particles. 
In all such cases from Eq.~(\ref{eq:P}) one can see $K^t$  remains finite at $r=0$. 
In the third case center just gets momentarily singular and
then there is a rebounce; all the shells reach $R = a(r)$ radius in
finite time and immediately bounce off. 
We get only one outgoing ORNG from the singularity 
and $K^t$ is always finite.

Now we calculate the redshift for these rays. Taking the source of radiation at naked singularity at $r=0$  and observer at some $r=r_o$,
which have four-velocities 
$
{u^a}_{(s)}=\delta^a_t$,  ${u^a}_{(o)}=\delta^a_t$, we get the redshift~\cite{FeliceClarke},
\be
1+z\propto \frac{(K^t)_s} {(K^t)_o}.
\ee
In the evaluation of the redshift, the behavior of the tangent
vector component $K^t$ is important. It is finite at the nonsingular observer. Due to the divergence
of $K^t$ for the infinite family, redshift  diverges for any non-singular observer. 

The observed intensity $I_p$ of a point source~\cite{DwiKan} is
$
I_p= \frac{P_0}{A_0 (1+z)^2},
$
where $P_0$ is the power radiated by the source into the solid angle
$\delta\Omega$, and $A_0$ is the area sustained by the rays at the observer.
The redshift factor $(1+z)^2$ appears because the power radiated is not
the same as power received by the observer. 
In case of ORNGs $A_0\propto R_0^2$ 
where $R_0$ is the area radius of the observer. 
As the redshift diverges and $A_0$ is finite, classically the
luminosity for all  null naked singularity along such families vanishes and no energy can reach an outside 
observer from the null singularities. 

Basically, for energy to reach an outside observer, the redshift should be finite for 
non-zero  duration. In this  case at the most it is finite along the first singular geodesic 
( it is just an instant) while it diverges very rapidly  for all  other singular geodesics. So the integrated energy reaching an observer from the singularity will be zero.
For the timelike naked singularities the luminosity function can be nonzero as the redshift is 
always finite and those can be  more dangerous for CCC.

To summarize, in this work, under very general conditions we have shown that, even in case, null naked singularities (geometrically) form in a spherical gravitational collapse, a non-singular observer can never receive any energy (and so also information) from them i.e. we cannot observe them.  
This means physically they are always (essentially) censored. 
This provides a strong support to CCC when  there is a continuous gravitational collapse without any bounce,
which will lead to a black hole or a null naked singularity. 
Though we have studied only ORNGs, the result can be easily 
generalized for nonradial null as well as timelike geodesics  as seen 
for the TBL models earlier~\cite{ShrirLum1, ShrirTime, NolanNrng} and we will 
get $K^t$ blowing up when a family of geodesics comes out.

For the case when only central shell gets singular and other shells eventually bounce off, a timelike naked singularity forms and the redshift is always finite for such a case. Hence if other factors do not play a role
one may be able to observe them.
Therefore the study of timelike naked singularities will play more important role in proving or disproving CCC. As such, it appears, only the central 
shell getting singular and rest of the shells bouncing off will need fine balance of factors i.e. it will need quite a bit of fine tuning and we would expect such timelike naked singularities will be rare.

Our main assumptions for getting the results are that the matter field is type I matter, the weak  energy condition holds and there is no shell crossing at least in the neighborhood of central singularity.  
Essentially except massless fields representing radiation in a single direction, all physically observed matter fields   fall under this category~\cite{HawkEllis}. Our assumptions are most natural and typically we want them to hold for any physically relevant case.   All the earlier known cases~\cite{HaradaICgc,TBLall1, PsjPrd93,
ShrirLum1, ShrirTime, NolanNrng,  Pfluid,  MagliTan2all,
HaradaTimeFam, Vaidya,   OtherMagliExact, PsjCmp94}
of naked singularities in spherically symmetric gravitational collapse of type I matter field fall under our analysis.
Our result is based 
only on local analysis of some Einstein equations and geodesic equations  near the central singularity.
We assumed that singularity forms in the gravitational collapse and then we analyzed them. If singularity does not form, then any way, we do not have to bother about their visibility and violation of cosmic censorship.
The other Einstein's equations will put further restrictions on various functions and may even avoid formation of singularities in some cases. But the conclusions drawn here will be always valid. 

Typically, even if shell crossing singularities form, extension of space-time through them is believed to be possible and they are not considered to be important for CCC. For us, as such,
even if they appear, as long as they are not in the neighborhood of the central singularity, our results still hold for the shell focusing singularity. Basically we just have to do our 
scaling after the last shell crossing happens near the singularity.

I thank Bala Iyer, Samuel Joseph, Sumati Surya and Madhavan Varadarajan for useful discussions.

\bibliography{ccf}

\end{document}